# A Network based Blockchain ecosystem for peer review publication

Omid FATAHI VALILAI

*Department of Mathematics & Logistics, Jacobs University Bremen, Bremen, Germany.*

*Campus Ring 1, 28759, Bremen, Germany*

E-mail address: *O.FatahiValilai@jacobs-University.de*

https://sites.google.com/view/fatahivalilai-omid/

**Abstract**

Blockchain as an emerging technology has transformed many traditional and conventional ecosystems and business models. Publication review system is one of the interesting subjects for transformation. The conventional models of publication review besides the urgent criticisms necessitates a transformation. This paper has investigated the most dominant shortcomings of literature review of publication review systems. Investigating the capabilities of Blockchain technology in the literature and its transformative capabilities, a network based Blockchain ecosystem is proposed. The ecosystem is providing a decentralized, transparent, and effective collaboration mechanisms for authors, reviewers, and other stakeholders in publication review processes. The main advantage of the ecosystem is committing the of all stakeholders and recognizing their contributions in a publication. Moreover, the ecosystem profits the stakeholders from the resulted reputation of the publication in future citations with its token model. The architecture of the proposed ecosystem supports the management of publication contents and sharing it for stakeholders. Moreover, it enables the configuration of smart contracts for governance of processes among stakeholders. It uses a token transaction Blockchain model to fulfill the token transactions. The capabilities and details of layers in the ecosystem has been discussed addressing the new paradigm of publication review ecosystem.

## 1. Introduction

Recently, the concept of Blockchain has gained the attention of academic researchers and professional business developers [1]. This is due to the transformative capabilities of Blockchain for globalized and decentralized collaboration [2, 3]. The application of blockchain has been elaborated in manufacturing and production industries [4–7], supply chain and logistics [8–12], medical fields [13], agriculture [14–16] and banking systems [16–18]. The wide domain of applications of Blockchain can be interpreted mostly considering its philosophy of decentralization and multi-organizational interoperability capabilities [19–23]. These capabilities are a necessity for nowadays global and wide world interactions and communications in different industries and academic environments [24–26]. This enables the Blockchain as a transformative tool for redefinition of conventional professional and academic processes.

One of the most dominant academic research processes can be defined as the academic publication [27–29]. Almost all researchers have dealt with this process in different forms like conference publications, Journal articles and Book publication [30]. This process is very important as it should secure the novelty of the research, contingency and integrity of the research, academic ethical necessities, copyright license concerns for distribution of publication and also proofing and editorial considerations [31–33]. Various platforms like journal publication or conference proceeding supporting solutions have been proposed to fulfill the main requirements of the review process [30, 33, 34]. The review process can be interpreted as an essential and integral part of consensus building which will lead to growth and development of scientific knowledge [35]. However, many criticisms have been raised for review process in recent years [36–38]. The majority of these criticisms can be interpreted due to the centralized mechanisms of review process at the publisher side and reviewers engagement in it [39–43].

This paper has proposed a network architecture Blockchain for publication review. Conducting a comprehensive literature review for publication review process, the main requirements for a decentralized review process will be discussed. With the aim of fulfilling these requirements, a new ecosystem in the form of a network architecture Blockchain will be discussed. The proposed network architecture Blockchain will be elaborated in terms of the architecture, the stakeholders and their engagement, the structure of the transactions, consensus mechanisms and policy of token distribution. The paper will discuss how the current publication business models can play a new role in proposed network architecture Blockchain ecosystem.

## 2. Literature review

The literature review will focus on two major perspectives:

- The current literature regarding the review process for publication and the related shortcomings and perspectives.
- The current literature about the recent achievements for using Blockchain as an emerging technology for transforming the conventional business models.

### 2.1. Publication and review process

Journal peer review can be interpreted as an insuring mechanism for an environment of unrestrained criticism for science research which makes it to work efficiently [28, 32, 38]. Peer review process has been recognized as a way to insure the democratic process of scientific publication [34, 36]. The institutionalization of review process has started after the second world war [35]. Peer review process was aimed mostly a quality assurance system [38]. The literature review in Table 1 has demonstrated that the review process is facing with major challenges which are mostly due to the decentralized and diverse standards for publication review. One of the most challenging issues are related to the not transparent communication channel between the reviewers and authors and the reveal of the review report for readers. On the other side, the role of publication and journal firms is complicated and not effective mostly in terms of efficient interactions with the reviewers [36]. Although, the reviewers who are academic/professional entities are mostly keen to participate in review process, still the motivational and appraisal mechanisms are not effectively designed [39]. although the reviewers are contributing in a research by giving valuable comments for improvements of the research, their efforts are not recognized in the published paper [44]. Also, the review ecosystem should be able to scale itself due to increase of quantitative submissions also being capable of adapting itself for new quantitative scaling due to expansion of research topics [45].

*Table 1. The review publication process in literature*

| Research | Focus of study | Criticisms for review process | Improvement requirements |
|---|---|---|---|
| Rennie [35] | Literature review of peer review evolution | - Unreliability<br>- Failing for validation of review<br>- Biased review<br>- Incompetent results | - A jury-based mechanism rather than a judge one<br>- Peer reviewers' compliment in the process<br>- The reassurance of readers |
| Mavrogenis et al. [38] | The requirements of peer review process | - Inconsistency of editors to address inappropriate reviews<br>- Review result with impossible requirements<br>- Discourages the authors | - Peer review should interfere in a positive way<br>- Providing respectful comments |
| Teixeira da Silva [46] | Challenges to open peer review | - Commercialization of science and information leading to biasing purely intellectual objectives with non-academic ones<br>- Wide variation in peer review quality between journals and publishers. | - Pre-publication peer review and post-publication peer review (PPPR)<br>- Providing a platform that accommodates for PPPR |
| Kelly et al. [36] | highlighting the pros and cons of peer review types | - Slowness of the publishing process<br>- Perceived bias by the editors and/or reviewers | - Suggesting the authors for fulfilling the shortcomings<br>- Accomplishing the review timely |
| Stern and O'Shea [39] | Proposal of future of scientific publishing | - Different journals having different distribution of articles<br>- keeping peer reviews confidential | - Enabling a publication first and then curating<br>- Emphasizing transparency<br>- Post publication appraisal |
| Hope et al. [44] | Proposing philosophy of scientific Peer review | - Incentivizing high quality peer reviews<br>- Insuring review quality<br>- Transparency of the peer review process | - Evaluation of the effectiveness of peer review<br>- Increase of the pool of well qualified reviewers<br>- Enabling the transparency of peer review process. |
| Reinhart and Schendzielorz [45] | Proposal of novel model for review | - Biased review<br>- Deficit model of peer review<br>- The effects of review process on domains like grant evaluation | - Adaptability and scalability of review<br>- High quality decisions in peer review |
| Parsons and Baglini [47] | Importance of neutral language in peer review | - Unprofessional peer reviews<br>- Peer reviewers with not language standards. | - Increased awareness for nonneutral language in peer review<br>- Explicatory and concreteness assertions |

As a conclusion from the literature review in this section, the paper highlights the following shortcomings in current peer review publication models:

- The performance of peer review process is not proper especially in terms of pace of review and quality of received feedbacks
- The contribution of the reviewers in a publication is not appraised effectively.
- The communication of stakeholders especially the authors and reviewers are not effectively established despite of various publication platforms

This paper has focused on shortcomings stated in the literature and is trying to provide the graph based Blockchain solution to overcome them. The idea is to enable a review ecosystem consist of mechanisms to:

- Increase the performance of peer review process by using effective and useful reviewers' feedbacks which help the authors to enhance their publications. This requires benefiting the reviewers beside the authors from the created contribution in a publication.
- Facilitate the communication of main stakeholders of the ecosystem; authors and reviewers for transparent and also mutually beneficial collaboration. This requires both, the scalable and rich pool of reviewers who are willing to conduct a high-quality review and also benefiting from the created contributions besides the authors.

## 2.2. Blockchain as an emerging technology

At the first stage, this section will give a brief introduction of Blockchain Technology. Blockchain is a distributed ledger mechanisms which is shared with a decentralized mechanism among stakeholders [5, 48]. The main benefits of blockchain can be stated in terms of its scalability to handle the transaction growth in a network effectively [13, 24], security [26], solving the deficiency of monopoly mechanisms [22]. Although the dominant application of Blockchain is Cryptocurrencies, in recent years application of blockchain based models have been considered in various business and industry models from Banking [18], agriculture [15, 26], manufacturing and production [2, 4], logistics [21, 49] and telecommunication [20, 50]. Most of the focus points for using the Blockchain are concentrated on definition of Transaction/Block, consensus mechanism, reward policy [2, 5].

At the second stage, the focus is to find the capabilities of Blockchain oriented solutions in different business sectors. The aim is to highlights the achieved merits and the fulfilled shortcomings in traditional and conventional business processes and scenarios. As a conclusion for the reviewed literature in Table 2:

- The capabilities of Blockchain for transparent meanwhile secured data and information communication is emphasized.
- The decentralized capabilities of Blockchain which enable the non-exclusive governance while saleable and adaptable processes is highly valuable.
- The consensus and smart contract mechanisms in the Blockchain which guarantee the validity and quality of generated contents inside the Blockchain while promotes and motivate the stakeholders effectively to work inside the ecosystem is highly referred.

*Table 2. The recent development in terms of Blockchain application in literature*

| Research | Focus of study | Blockchain characteristics | Dominant advantages |
|---|---|---|---|
| Radmanesh et al. [51] Aghamohammazade and Valilai [2] Ahmadi et al. [3] | - Cloud Manufacturing and Production system<br>- Cloud Supply network | - Service matchings builds the transactions and blocks<br>- Miners are service Matchers<br>- Reward function is based on the optimization effort<br>- Consensus mechanism to insure feasibility of service matching | - Service composition among providers and demanders are managed in decentralized mode<br>- The optimization as an NP-hard problem is more efficiently managed<br>- New framework for data sharing among firms in supply network is established |
| Lee et al. [6] | - Cyber-Physical Production Systems integration and synchronization | - Fulfilling the security and privacy by using advanced cryptographic algorithms<br>- Enabling the distributed storage of data<br>- Fault free with global consensus mechanism | - Interoperability and integration from a global perspective<br>- Mapping the raw data to information and knowledge for network nodes<br>- Enabling rapid decision-making mechanism |
| Prashar et al. [26] | - Food safety through traceability of agriculture products | - Smart contract among the producers<br>- The store of traceability data inside the Blockchain<br>- Validation of data through consensus mechanism | - Traceability through whole chain of food production<br>- Optimizing the performance of production system<br>- Effective and fast data communication |
| Osmani et al. [17] | - Next generation of banking and finance | - Cryptocurrency style of financial transaction support | - Cost reduction for financial infrastructure<br>- Cost reduction for transaction fulfillment<br>- Privacy and transparency<br>- Scalability and interoperability |
| Mohamed and Al-Jaroodi [52] | - Smart manufacturing for Industry 4.0 applications | - Resource matching through transaction and clock definition<br>- Smart contract mechanism to handle the financial transaction | - Increasing trustable and secure data communication<br>- Enabling the data communication among different resources working with a XaaS approach |
| Hwang et al. [53] | - Transparency for energy prosumers | - Cryptocurrency style of energy trading | - Convenient and safe data collection<br>- More efficient Transaction model<br>- Connection of various energy sources and users |

| Research | Focus of study | Blockchain characteristics | Dominant advantages |
|---|---|---|---|
| Yuan and Wang [54] | - Intelligent transportation system | - Transportation data in form of transactions and blocks<br>- Smart contract mechanism to handle the financial transaction | - Enabling the security and trust in data communication<br>- Using the capabilities of distributed computing<br>- Crowdsourcing enabled in a decentralized architecture |
| Turkanović et al. [55] | - Higher education credit platform | - Students evaluation records in form of transaction and blocks<br>- Proof of concept as consensus mechanism | - Trusted and privacy enabled credit storage and sharing<br>- Distributed verification mechanism for credit evaluation |
| Wang et al. [56] | - E-Healthcare system | - Patient biomedical data storage through transaction and blocks | - Distributed remote physical conduction monitoring<br>- Trusted and privacy enabled credit storage and sharing<br>- Low power consumption |

The literature has illustrated the transformative capabilities of Blockchain technology as an emerging technology. Especially, the decentralization of processes together with consensus mechanisms can help ecosystems to benefit the expandability and scalability of governance processes without being restricted in centralized exclusive authority. Moreover, the other technologies and paradigms evolving inside Blockchain technology like Web 3.o will be great opportunity to enable the interoperability and post processing of data inside the Blockchain more effectively to support the ecosystem interactions [57–59].

## 3. Network Blockchain for peer review publication

### 3.1. The problem definition and solution perspectives

Considering the two perspectives investigated in the literature review, the paper proposes a solution to use the Blockchain capabilities to create a new ecosystem for publication review. The main concentration in this ecosystem will be the increase of review process performance in terms of quality and speed of review. Also, the contribution of all stakeholders inside this ecosystem on the resulted publication should be insured. This includes the author(s), reviewers and all the parties which are adding value to resulted contribution in the publication. The ecosystem should be saleable, adaptable and also it should enable the post processing of the contents in terms of providing the accessibility of contents effectively and benefiting the contributing stakeholders from the quality of the paper publication. Also, one of the most important features will be the effective content management in the terms of availability and post processing for being used by other researchers in their research studies.

To fulfill the above problem statement, the main idea is to use the capabilities of Blockchain to shape the publication review ecosystem. In this ecosystem, the stakeholders shall interact in terms of Blockchain network structure, the interaction model in terms of how the publication can be broadcasted and reviewed effectively with the designed consensus mechanisms. Also, the smart contract will enable the fair and effective benefit sharing among all stakeholders contributing to a publication. The decentralized nature of the Blockchain will insure the adaptability and scalability of the ecosystem meanwhile insure the wide access to the contents. Using the features of Web 3.0 in this ecosystem, a high potential for pos processing of the publication will be created which will address the initiation and development of conventional and transformative business models connected to the ecosystem.

It worth mentioning that the closest paradigm to the proposed solution in terms of ecosystem architecture can be regarded the Non-Fungible Tokens (NFTs). NFT has been offered as a paradigm for digital asset trading [60–62]. Using the mechanisms of smart contract on Blockchain based cryptocurrency models, NFTs are used to transmit the ownership of digital images and media [63, 64]. However, the prosed solution in this paper, will be fundamentally distinctive as the content of the publishing materials are going to be processed and reviewed for better contribution to literature. The NFTs are relying on the value of the content more from side of the trader side [65], however in a publication the quality of the published material itself will be assessed through a separate consensus mechanism which would increase the contribution value by the contribution of the reviewers.

### 3.2. Network Blockchain solution for peer review publication: conceptual model

The overall conceptual model of proposed network based Blockchain framework has been illustrated in Figure 1. The ecosystem transactions are managed through these three layers in which especial processes are designed in terms of publication submission, review and contribution management through token transactions among stakeholders. The Framework is composed of three layers which are interacting with each other:

- The "Publication Graph" layer which is the highest layer in the framework. This layer contains all the publication data consisting manuscript contents, review reports and meta data regarding the authors, reviewers and other stakeholders' contribution. This layer shapes the network of publication and their dependencies in terms of the citations relating them together. These relations will trigger the smart contracts in the "Smart Contract Layer" and establish the transactions for transfer of the tokens through the smart contract stakeholders. Also, this layer is equipped with Blockchain API to establish the communication of third-party value-added service providers to the ecosystem.

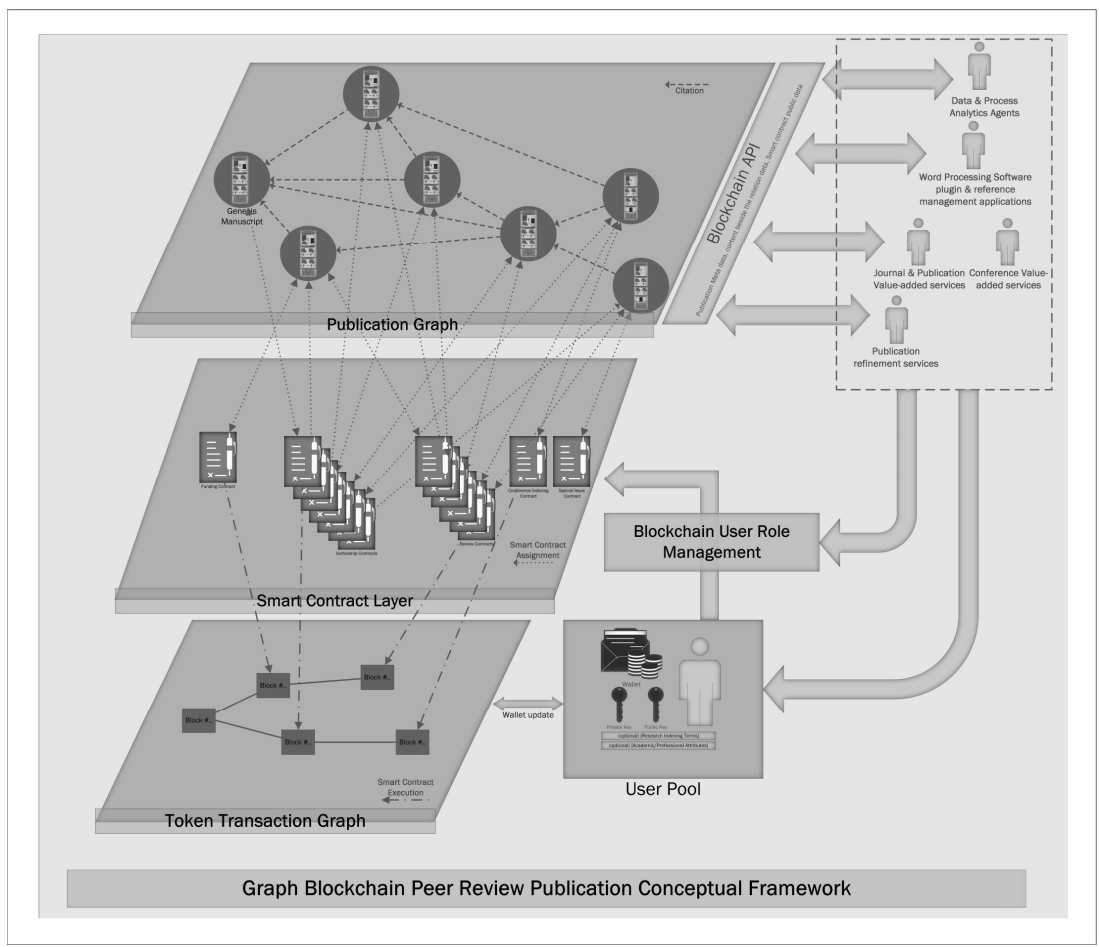

*Figure 1. The Conceptual model of graph based Blockchain Peer review Framework*

- The "Smart Contract Layer" which is responsible for establishment of contractual relations among different stakeholders in the publication review ecosystem. These relations can be categorized in terms of authorship agreement for specifying the share of the contributions among the authors, the reviewed contribution in terms of adding value and guiding the authors for better proposing their publication, the research grants announced and assigned to the contribution of the authors, the special issue of a journal or a conference indexing contract including specific papers in terms of topics and contribution or any other value added which is created via third parties through API connections. All the token transactions inside the ecosystem are trigged via the smart contracts inside this layer. The stakeholders can contribute ad play role inside the ecosystem by establishing a smart contract instances from the provided templates inside this layer .

- The "Token Transaction Graph" which fulfills the token transaction in ledger. This graph is the connections of blocks consisting of the token transactions among the stakeholders. As stated earlier, the blocks are triggered by smart contracts and transmit the token from an address in the ecosystem to another address.

Besides the three layers, the framework consists of user pool which manage the users interacting inside the ecosystem, "Blockchain User Role Management" section which manages the users' behaviors and interactions inside the ecosystem in terms of their engagements in „Smart Contract Layer" and finally the Blockchain API which enables the third-party interactions with the ecosystem. the details of functionalities inside each section is described in details in proceeding sections.

### 3.2.1. User pool

The user Pool will act as the user profile management section in the ecosystem. As illustrated in Figure 2, the stakeholders should establish their accounts in the ecosystem. The accounts will contain two major attribute categories. The first category is keeping the token and wallet private information like the public and private key information besides user preferences for their wallets. This category is connected to the "Token Transaction Graph" and is updated when the transactions are implemented inside the "Token Transaction Graph" like receiving of token and their withdrawal. The second category of data are related to optional research/academic and professional attributes. Stakeholders like researchers can enrich the user pool with data and attributes regarding their research expertise, key words, scholar ids like google Scholar, ORCID and other professional social media attributes. These attributes are important as they can influence the performance of mechanisms in the "Blockchain Role Management Section".

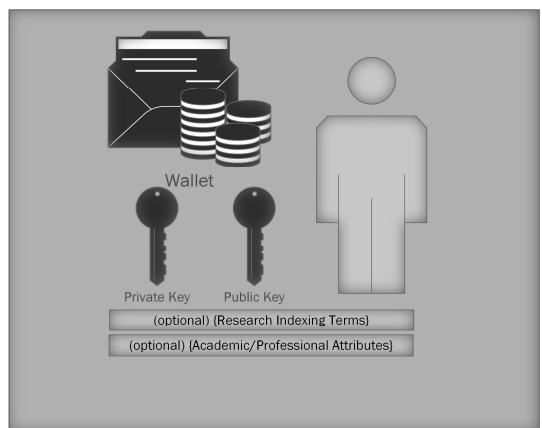

*Figure 2. User Pool overall structure*

For the stakeholders from Professional domain like journals, conferences or any other third service providers the second category of data and attributes would describe their business and marketing data and capabilities which will help the authors to find and interact more efficiently to improve the publication qualities. Examples of these category stakeholders are journal editorial services, conference committee board, funding agencies and proofing service providers. The details of collaboration types are discussed in section "Blockchain User Role Management".

### 3.2.2. Smart Contract Layer

This layer facilitates the initiation and execution of smart contracts in the ecosystem. the users from the user pools via the "Blockchain User Role Management" collaborate with the other stakeholders in this layer. The transactions regarding the token transmissions among the users in "Token Transaction Graph" will be triggered from the smart contracts in this layer as shown in Figure 3.

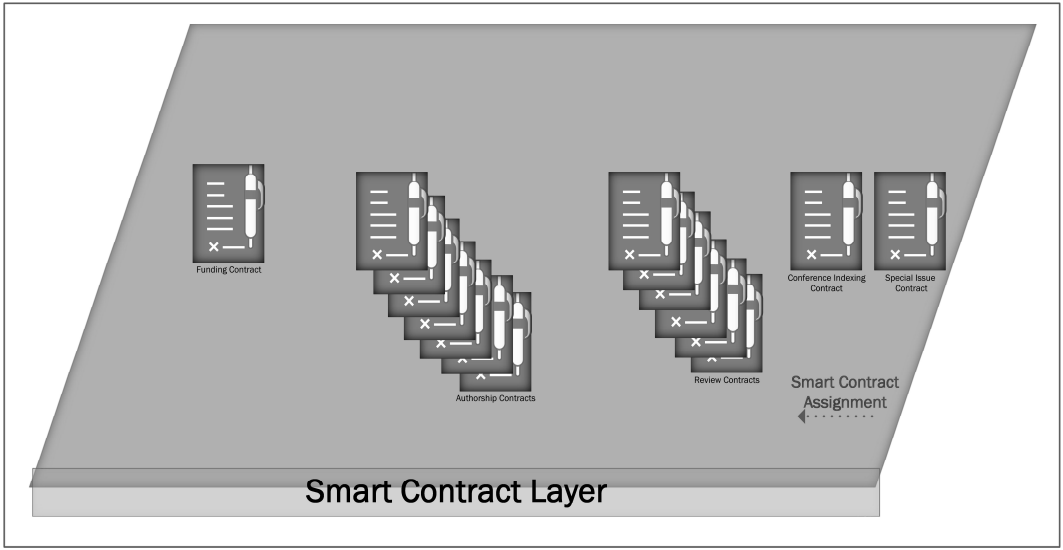

*Figure 3. Smart Contract Layer illustration*

The process of signing the contract will be handled by "Blockchain User Role Management" section. The major types of contracts are:

- Authorship contracts: If an author or a group of authors intend to submit a manuscript inside the ecosystem, they should sign a contract of authorship in this layer. The smart contract will initiate one node in "Publication Graph" based on the authorship smart contract signed with the authors. This type of contracts is executed in different occasions. They trigger the token withdraw from authors' wallet when the manuscript is confirmed and signed by the required number of the reviewers. Also, when the published manuscript receives citations from other confirmed publications, the authorship portion of received token from those manuscripts will be distributed among the authors based on the authorship smart contract structure.
- Review contracts: when a reviewer accepts the review assignment of a manuscript, the reviewer would be entitled to sign a smart contract. This contract will withdraw the tokens from the wallet of reviewers if they sign and agree to review the manuscript. When the confirmed manuscript receives the citation, the review portion of received token from those manuscripts will be distributed among the reviewers based on the review smart contract structure. When a manuscript receives enough confirmations by the reviewers it would be treated as an accepted publication in ecostsem.
- Funding contracts: when a funding agency intends to announce a funding in a special topic or intends to assign a specific fund to specific authors, this type of contracts can be used. The token from the funding agency would be transferred to authors based on the specific terms in these smart contracts. The funding contracts can change the authorship contracts in terms of transfer of token (or a portion of tokens) from funding agencies instead of authors wallet. Also. In case of receiving token after the confirmation of publication in ecosystem and receiving citations, special terms for receiving of token (or a portion of tokens) can be established which transfer the token to funding agency wallet.
- Journal indexing/Special issue/book publishing/Conference indexing contract: the ecosystem is enabling the varieties of users like book/journal publishing firms or conference organization committees to have the review and publishing of their manuscripts in the "Publication Graph".

>This type of contracts will establish special terms via the "Blockchain User Role Management" for the initiation and execution of authorship and review contracts. The ecosystem enables the payment of conference fees to be fulfilled via the citations the paper receives in future from citing publications. The review contract can also be tailored based on specific requirements of this type of contracts.

The smart contract can be extended to encompass various types of value-added services especially considering the communication via Blockchain API. For example, the Proof-reading service providers can have their smart contract enabled in this layer via "Blockchain User Role Management" section. This will enable them to conduct the proof-reading services and invest their tokens in a manuscript while agreeing with the authors to benefit from the received token in future citations. As observed, all the stakeholders which are contributing to a manuscript from authors. Reviewers, funding agencies, conference organization boards and proofreaders should invest their tokens accordingly in related smart contracts. This means all stakeholders are investing in a manuscript and after the confirmation and receiving the citations from the other manuscripts they will receive tokens based on their smart contracts and the reputation of the publication to receive citations. This is one the efficient mechanisms inside the ecosystem which promotes the quality while recognizing the contribution of all stakeholders and engaging them in terms of investing their tokens and commit to the success of the publications.

### 3.2.3. Publication Graph

This layer contains the manuscript contents and the related meta data regarding the contributions which are conducted for each manuscript. This layer forms a graph in terms of each manuscript acting as a node and the citations in terms of the vectors (the vector starts from the citing manuscripts toward the cited reference publication). This graph will also be in relation with the "Smart Contract Layer". To better understand the logic behind this layer, the paper first describes the structure of each manuscript which would be a node in "Publication Graph" layer.  The manuscripts will have four main components as illustrated in Figure 4. All the information regarding these components are stored in "Publication Graph" layer.

- The first component is authorship. This component is shaped by the help of authorship smart contract. When author/authors finalize their author ship smart contract, the related meta data regarding their attributes besides their manuscript data is updated in this component. As stated earlier through the agreed terms in authorship smart contract the authors will put definite number of tokens in their manuscript which would be blocked in the manuscript. After a manuscript receives the confirmation from reviewer(s), these tokens are sent to all the citations inside the manuscript based on the Blockchain policy. If authors refuse to continue, the reserved tokens in the manuscript will be processes by the authorship policy ranging from completely/partially being refunded to authors or being used inside the ecosystem for performance refinement.
- The second component is the confirmation component. This component contains all the conformation signed by the reviewers in the ecosystem. As stated earlier, the reviewers are assigned by the "Blockchain User Role Management" section in ecosystem. If a reviewer accepts the review of a manuscript, then the review smart contract is signed, and it would be recorded in this component inside the manuscript. It worth mentioning that in case the reviewers find merits in a manuscript after signing the smart review contract and investing tokens, they will write their review report including their refinement suggestions which would be in this component besides their tokens. In case the reviewer signs and confirms an article (accept the article), the acceptance report is recorded besides the current versions of the manuscript contents in this component. This will allow the transparent communication of revision and the review process not only with authors but also with all the future audiences in ecosystem. As the reviewers are contributing in terms of publication, they are awarded in future based on the citations (through the tokens assigned by citing manuscript) which the publication receives. The number of required conformations for a manuscript depends on the policy of ecosystem and can be configured in "Publication Graph" layer. This can be interpreted as a consensus mechanism in "Publication Graph" layer. The paper calls this consensus mechanism Proof of Review (PoR) which confirms and accepts the manuscript nodes in the ecosystem based on the signatures and confirmations they receive from the reviewers. This demonstrates the decentralized nature of the ecosystem in which the reviewers are collaborating with the designed PoR consensus mechanism to give credit and validate the manuscript in "publication graph" layer.

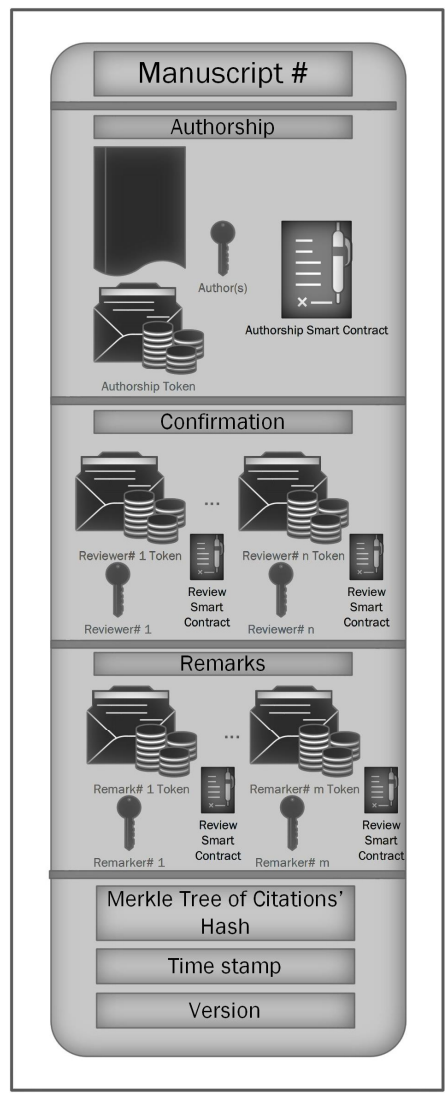

*Figure 4. The manuscript components*

- The third component is the remarks component. This component contains all the other related smart contract data and attributes related to the manuscript which can be related to funding agencies, proof reading services, special issue journals, book publication and conference events. As stated earlier, all the remarks will store token in this section which will be released as the manuscript is signed and confirmed with required number of confirmations and will be share among all the cited manuscripts in "Publication Graph".
- The last component will be the meta data for the manuscript node which are version, time stamp and Merkel Tree of citations' hash. These data improve the performance of analyzing the "Publication Graph" information. Each manuscript will have a unique hash which is built by the definite function inside the smart contracts using the private keys of authors and reviewers and remarking agents.

The paper considers this article as the "Genesis Manuscript". The "Genesis Manuscript" is the first block in "Publication Graph" layer which would not transfer any token to any other manuscript as it would be the first node initiated in "Publication Graph" layer. The first relations inside "Publication Graph" layer will be established in terms of new nodes (manuscripts) citing the "Genesis Manuscript" as illustrated in Figure 5. When a manuscript is initiated inside this layer, the manuscript node with authorship will be established and the manuscript hash will be built up from the authorship public and private keys signed the authorship smart contract. As the paper is reviewed, the second component is formed through the signatures of the reviewers via their smart contracts and update the hash through encapsulation the private keys of the reviewers. A reviewer may multiple times review and revise a manuscript which means the second component will be updated each time resulting in new hash. Until the manuscript receives the enough number of the confirmation by the reviewers the first component can be updated by the authorship smart contract and the third component by the remaking agents. When the enough number of the confirmations are received the manuscript, node is locked and all the related smart contracts are triggered and would be locked to prevent new modifications.

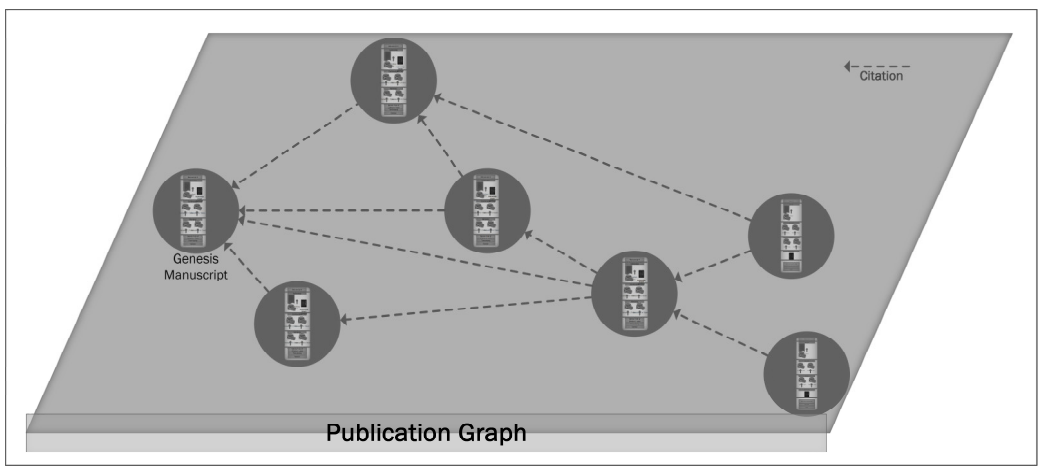

*Figure 5. The Publication Graph*

As mentioned earlier, the vectors inside the "Publication Graph" layer are connecting manuscript based on the citations created inside the ecosystem. The ecosystem will use the features of third-party service providers through the "API Blockchain" section to facilitate the manuscript preparation in terms of generating citations. The hash of the manuscript will act like the Digital Object Identifier (DOI) and enable the unique referencing and establishing the correct relations between the manuscripts. After a manuscript receives enough confirmations, the citations will be used to trigger the related smart contracts inside the manuscript and transmit the tokens through the ecosystem.

### 3.2.4. Blockchain User Role Management

As described thought the former sections, all the stakeholders inside the ecosystem should establish their user profiles in user pool. This encompasses authors, reviewers, journal publishers, conference committee boards and value-added service providers. As all the transactions inside the ecosystem should be conducted through smart contracts, the "Blockchain User Role Management" has the responsibility to define, configure and establish the proper types of smart contracts for users besides providing required interfaces for effective interactions. This section can be modified through evaluation time of ecosystem to improve the efficiency of the processes and adjusting the behavior model of smart contracts. Also, any new business process should be embedded in this section to connect the users to proper types of smart contracts. The required fundamental types of mechanism in this section are:

- The authorship mechanism: this mechanism is enabled when author/authors intend to publish their manuscript in the ecosystem. It would facilitate the collaboration of the authors together to sign their authorship smart contract. If funding agencies should be engaged, this section will initiate their role in funding smart contracts and binds it to the manuscript which would be then a node in "Publication Graph" layer. As there would usually different revisions which may encounter the change of the authorships and their contribution levels, again this section will handle the modifications of the authorship contracts. As an example, the configuration regarding to limiting the token receipt by self-citations can be handled in this mechanism.

- The reviewer mechanism: this mechanism assigns the manuscripts listed in "Publication Graph" layer to the reviewers in the ecosystem. Just like all the other review mechanisms in traditional publication mode, the engagement of reviewers to accomplish the review is required. However, this mechanism is fulfilling the former shortcomings. The ecosystem uses a decentralized mechanism which select random users (which have agreed in their user profile creation to contribute as a reviewer) to accept and sign the review smart contracts. To increase the effectiveness of this selection, the mechanism will use the historical authorship smart contracts of the users, their key words and indexes in their profile creation and their former review contributions. The reviewers can deny their invitations for signing the review smart contracts. If they sign the review smart contract, they should transfer the committed token to the manuscript. The paper calls this the proof of review (PoR) model which resembles a kind of Proof of stake model. This is as stated will insure both the contribution of the reviewers to increase the quality as the ecosystem is considering the review as a value-added role and will share the received tokens from citations in future for the reviewers. The reviewers are assigning tokens to the manuscript committing to its success as the manuscript was found to be worth of reviewing. For special type of contracts like special issues and conference organizations, the specific methods for review selection and token reserving would be possible. The review mechanism will continue to select and propose the review smart contracts to the reviewers until the required confirmations are received or the authors refrain from publication progress.
- Book/Special issue/Conference publication mechanisms: these mechanisms enable the special types of manuscript publishing in the ecosystem. the principles of authorship and review smart contracts are maintained but the policies for token transmissions and special criteria for reviewer selection maybe applied. This enables overriding for a portion of conference registrations or open access reviews as the required processes are fulfilled in the ecosystem and the revenue business model can be fulfilled for stakeholders from the reputation of the resulted publication in future.

### 3.2.5. Token Transaction Graph

This layer will accomplish the fulfillment of token transactions inside the ecosystem. while the "Publication Graph" is supporting the manuscript content and meta data for authorship/review and value-added activities for the publications, all the required transactions for transferring token among the users inside the ecosystem will be accomplished in this layer. The ecosystem can use different types of Blockchain token architecture for this layer. The important requirement is the support of smart contract mechanisms in "Smart Contract Layer". The token model in this layer is suggested to be a graph-based token model as illustrated in Figure 6, but also the conventional chain shaped models are also applicable.

The proposed framework in this paper is enabling a plug-in architecture for handling "Token Transaction Graph" layer. It means the "Smart Contract Layer" can be deployed on different Blockchain token models to support the "Token Transaction Graph". As the paper is considering a special token distribution, it would be better to offer a ground and self-created token model based on pre-mined tokens. As the user creates their user profile in the pool, they will be assigned the private/public keys. These keys are used to secure their wallets and tokens inside the ecosystem and also for signing the smart contracts and creating hashes for the manuscript published in ecosystem. The Ecosystem promotes the researchers to establish their user profiles inside the ecosystem and gives gift tokens based on the researchers' academic profiles. Moreover, the tokens can be purchased from markets through the crypto exchanges when the token is listed. As the ecosystem evolves, the value of tokens will increase which will result in increase of the value of user wallets asset. The ecosystem also will donate tokens to promote and support especial research trends in form of funding smart contracts. The rate of token cycling can be controlled through the "Blockchain User Role management" section and smart contracts. The general policy will be the configuration to enable the receipt of more tokens than spent tokens of authors and reviewers for high quality publications.

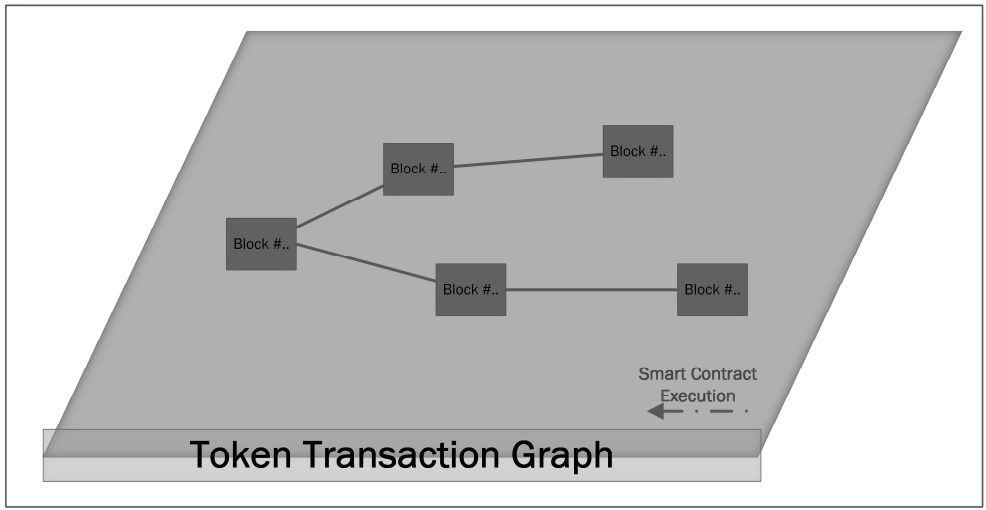

*Figure 6. The Token Transaction Graph*

### 3.2.6. Blockchain API

The Blockchain API is designed to enable the ecosystem manuscript content communication with value-added service providers. These service providers can have either user role inside the ecosystem and be engaged through the "Blockchain User Role Management" section in smart contracts or they can only provide services to increase the effectiveness of ecosystem processes. Especially with the application of Web 3.0, the contents of the "Publication Graph" layer can be effectively processed and used to create valuable insights. The current designed APIs can facilitate the following value-added services in ecosystem:

- Word processing and reference management applications which will use mostly the manuscript content data in the "Production Graph" layer. These applications facilitate the effective search and retrieve of publication in the ecosystem and indexing them for ecosystem users. Although the ecosystem as a distributed ledger can be downloaded and updated in every user's computer, the search and query for finding the proper publications would be hard and both storage space demanding. The reference management applications help the users to have access to the publication data effectively in terms of finding relevant publications and moreover citing them in their manuscript. Nowadays, word processing plug-ins facilitate the citation and reference list creation for authors. The same capabilities will be provided for the users to embed the citations and generate the reference lists based on different reference styles and article templates.
- Book and journal publishers/conference value added services which as described are users in the ecosystem. As the ecosystem is acting as a distributed ledger and provides the access to publication data and contents, these stakeholders can use it as their publication repositories and develop their presentation user interfaces for special designed publication directories. This can resemble a conference proceeding for historical archive of a conference events or special book/journal series.
- Data and process analytics agents which will use the publication content data in the ecosystem to generate valuable insights and knowledge. The generated knowledge can be used to demonstrate the research trends, subject funding trends, biometric researcher characteristics and even the effective collaboration patterns for researchers. In fact, many data analytics solutions can use the "Publication Graph" layer data as their source for knowledge extraction.
- Publication refinement services which will be the value-added service of journals' editorial boards to the ecosystem users. This can provide the users based on their demand for special types of smart contracts in which the editorial boards will provide advisories for the authors to help them fulfill the reviewers' comments and receive the confirmation more efficiently. It worth mentioning that would require the publication refinement services to invest their tokens in manuscript and will receive tokens in return in future publication citations in the ecosystem according to quality and reputation of the publication. The ecosystem is transforming the roles on editorial boards from judges to advisors which help the authors and reviewers based on their knowledge and expertise.

The Blockchain API can be extended with new value-added processes and functionalities which promotes business model development in publication review domain.

## 4. Discussions and Conclusions

In this paper, a new ecosystem for publication review has been proposed. Considering the shortcomings of current state of the art publication review models and capabilities of Blockchain technology, a new distributed and decentralized framework has been designed. The main capabilities can be summarized as:

- The performance of this ecosystem in comparison with the traditional review models would be incomparable. The ecosystem user pool provides the access to a very great repository of willing reviewers which are committed not for judging the manuscript but for collaborating and advising the authors. The model is effective as it is not promoting a criticism-based approach for review but a collaborative approach. Meanwhile, the contribution of all stakeholders ranging from authors and reviewers are fulfilled. The token model inside the ecosystem engages the stakeholders in investing in manuscripts and benefiting from the publication quality in terms of receiving citations and the publication reputation. The ecosystem is completely scalable due to its decentralized architecture.
- The communication of authors and reviewers is increased and refined in terms of transparency and integrity. The centralized solution through different journals and their editorial boards has increased the complexity and criticisms of decisions for authors. Challenges of selecting the proper journals for authors and transferring their manuscripts from one journal to another one has been fulfilled in this platform. The ecosystem is decentralized, and the role of journals has changed and transformed greatly. The journals' editorial boards would help the manuscript refinements and advising the authors rather than judging and making the probable criticized bias decisions. The access to publication content data is provided for all stakeholders and provide the revenue models would be transformed to value-added services which would really help the authors for effective and high-quality publication.

The paper also discusses some concerns observed in publication review literature which can be approached effectively by the capabilities of the proposed ecosystem:

- Considering the recent arguments about the blind peer review or non-blind review processes, the ecosystem is designed mainly with concentration of on non-blind review process as it is believed the role of reviewers has been transformed to facilitators and contributor collaborators in the ecosystem. However, relying on capabilities of Blockchains to provide secure and anonymous transactions, the framework can enable the blind review system which will not reveal the authors' and reviewers' identity until the required conformations are obtained.

- There are also some concerns about the non-ethical behaviors of authors and reviewers which the ecosystem has the capabilities to fulfill them with proper configuration of smart contracts. As an example, the self-citations have been discussed both positively and negatively. The ecosystem can use the smart contract mechanism to control the trend. The ecosystem by default is not against the self-citations, however, if a journal is against this characteristic in research, it can use the Blockchain API for configuring specific smart authorship smart contracts which eliminate the token transmissions to authors in a publication cited with the publication of same authors.
- Recently, there has been a trend for shifting to open access style of publication. The author considers this trend as a strategic move of publishing businesses to secure their revenue model to receive the publication fee from authors rather than the audiences demanding to access them. The proposed ecosystem is drastically transforming the paradigm here. The token model in this ecosystem will generate wealth for all stakeholders and distribute it among them. Meanwhile a worldwide and global free access to publication data will be provided for all researchers through the "Publication Graph" in the ecosystem.

The ecosystem can also be affected by the configuration of smart contracts. This is an important and interesting topic for future studies and developments. Considering the speed of token cycling and stakeholders' preferences in their smart contracts, the equilibrium status analysis using the game theory methods and system dynamics is highly proposed. The author believes that ecosystem will find its balance and can be configured wisely by the consensus of stakeholders for smart contract definition.

## 5. Acknowledgement

The idea has been supported in terms of documentation in its early stages of initiation with the help of Mr. Ehsan Aghamohammadzade (ORCID:0000-0003-4560-4265). The author appreciates the help.